\documentclass[12pt]{elsart}

\usepackage{subfigure}

\usepackage[dvips]{graphicx,color}
\usepackage{dcolumn}
\usepackage{bm}
\usepackage{ulem}
\usepackage{soul}

\begin{document}

\begin{frontmatter}
\title{A comparative study of angle dependent
magnetoresistance in [001] and [110] La$_{2 / 3}$Sr$_{1 /
3}$MnO$_{3}$}

\author{Soumen Mandal\corauthref{cor1}}
\address{Department of Physics, Indian Institute of Technology
Kanpur, Kanpur - 208016, India} \ead{soumen.mandal@gmail.com}
\corauth[cor1]{Corresponding Author. Present address: Institut
N\'eel, Grenoble, France}

\begin{abstract}The angle dependent magnetoresistance study on [001]
and [110] La$_{2 / 3}$Sr$_{1 / 3}$MnO$_{3}$ thin films show that
the anisotropy energy of [110] films is higher when compared with
a [001] oriented La$_{2 / 3}$Sr$_{1 / 3}$MnO$_{3}$ film of similar
thickness. The data has been analyzed in the light of multidomain
model and it is seen that this model correctly explains the
observed behavior.\end{abstract}

 \end{frontmatter}

\maketitle

\section{Introduction}
Double exchange pervoskite manganites, such as La$_{2 / 3}$Sr$_{1 /
3}$MnO$_{3}$ (LSMO), show colossal magnetoresistance\cite{Fork}
and have $\sim$100{\%} spin polarization\cite{Bowen,Tokura1}.
Hence, this material forms an excellent candidate for studying
spin polarized quasi particle injection in various types of
systems. Secondly, the fact that magnetoresistive materials have
immense technological application, it is essential to understand
the underlying principles of the plethora of extraordinary
phenomena seen in this class of materials. The exact microscopic
mechanism for very large magnetoresistance(MR) is not very well
understood in these systems.

In this paper a careful study of magnetoresistance as a function
of angle between applied field and current direction in [001] and
[110] epitaxial films of La$_{2 / 3}$Sr$_{1 / 3}$MnO$_{3}$ is
reported. In both the cases MR at 10K and 300K was measured. While
hysteresis in resistivity at 10K is seen for both films, at 300K
only [110] films show hysteresis. This clearly demonstrates that
the anisotropy energy of [110] films are larger than that of [001]
films. Arguments and explanation of the observed data is based on
the multidomain model given by O'Donnell
\textit{etal}\cite{Ecksteinprb2}.

\section{Experimental Details} Thin epitaxial films of La$_{2 / 3}$Sr$_{1 / 3}$MnO$_{3}$
were deposited on [110] and [001] oriented SrTiO$_3$(STO)
substrates using a multitarget pulsed excimer laser [KrF,
$\lambda$ = 248 nm] ablation technique. The deposition temperature
(T$_{d}$), oxygen partial pressure pO$_{2}$, energy density
(E$_{d}$) and growth rate (G$_{r}$) used for the growth of 150 nm
thick layers were, 750$^{0}$ C, 0.4 mbar, $\sim$2J/cm$^2$ and
1.3{\AA}/sec respectively. To fully oxygenate the samples, the
deposition chamber was filled with O$_2$ to atmospheric pressure
immediately after the growth and then the sample was cooled from
750$^{0}$ C to room temperature. Epitaxial growth in two sets of
films with [110] and [001] directions normal to the plane of the
film was established with X-ray diffraction measurements performed
in the $\theta - 2\theta$ geometry. The measurements of
resistivity as a function of temperature, magnetic field strength
and the angle between field and current were done using a 4.2K
close cycle refrigerator with a fully automated home made setup
for applying fields at varying angles between 0 and 2$\pi$ with
respect to the direction of current\cite{Patnaik}. The sample was
mounted in way to keep the field in the plane of the sample for
all angles of field. The room temperature resistivity of these
samples is in the range of $\sim 2.7 - 2.8$ m$\Omega - $cm.
Isothermal hysteresis loops were measured for both the samples
using a commercial magnetometer (Quantum Design MPMS XL5 SQUID) by
applying field at various angles in the plane of the film. For the
measurement of magnetoresistance in four probe configuration films
were patterned in the form of a $1000 \times 100 {\mu}m^2$ bridge
with photolithography and wet etching such that the long axis of
the bridge was parallel to [001] and [100] direction for the [110]
and [001] oriented films respectively. The [001] and [110] films
were first characterized using a SQUID magnetometer to determine
the easy axis . The data for such measurements have been published
elsewhere\cite{Easy}.

\section{Results and Discussions} 
\begin{figure}[t]
\centerline{\includegraphics[width=10cm]{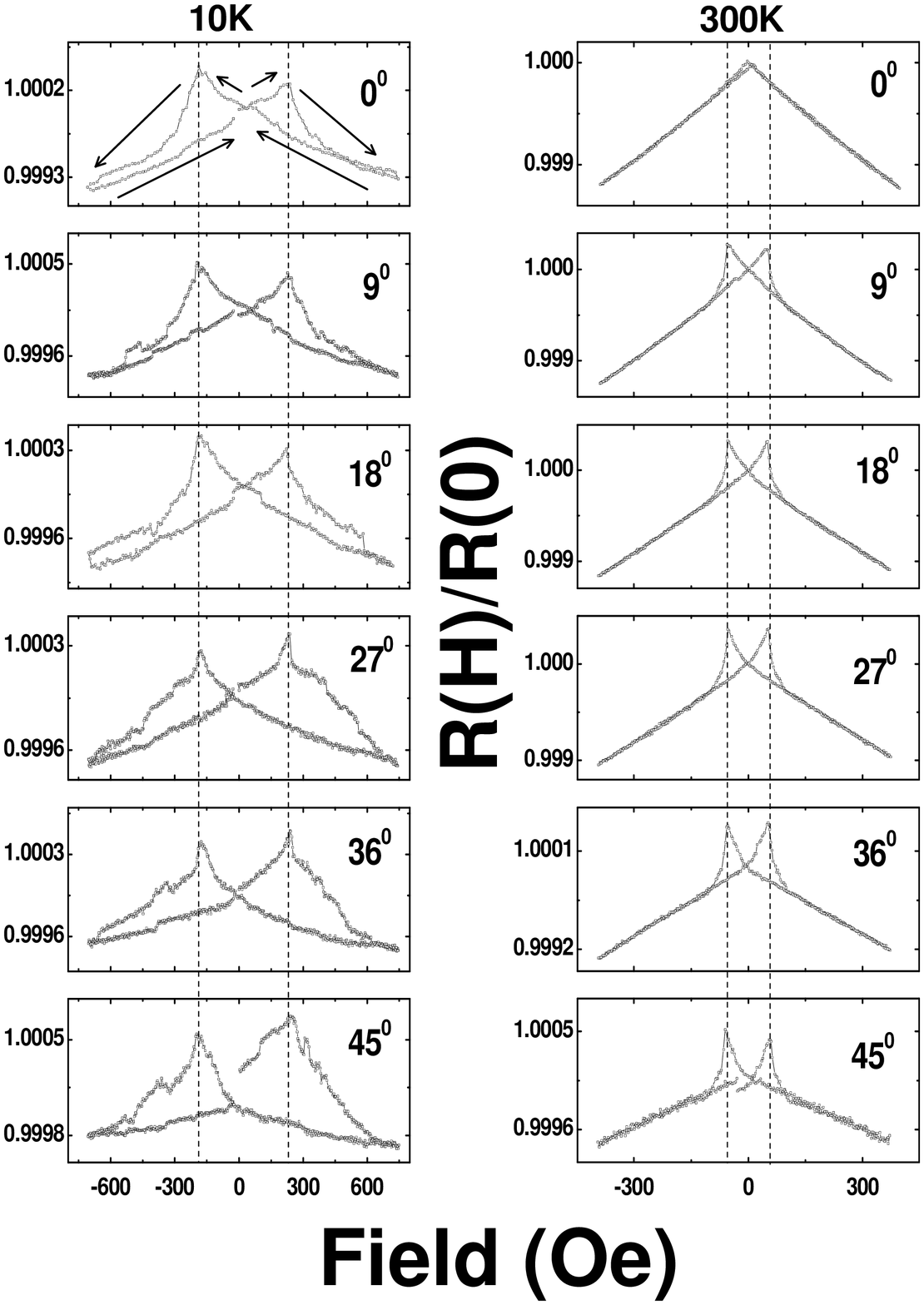}}
\caption{R(H)/R(0) vs. H data for the [110]  sample at 10 K (left
panel) and 300 K (right panel) taken at various angles between the
applied field and the current direction. The current in this case
is flowing along the easy axis, and the hard axis is in-plane and
90$^0$ to the current direction. Arrows in the figure mark the
trajectory followed by the resistance as the field is swept from
positive or negative extremities. At 10 K the resistance increases
superlinearly as the field reduced from 700 Oe till it reaches a
critical negative value H$_c$. On increasing the field further in
the reverse direction, the resistance drop, first rapidly and then
gradually. The resistance profile during -H$_{max}$ to +H$_{max}$
field sweep is a mirror image of the +H$_{max}$ to -H$_{max}$
sweep. A large hysteresis is evident in the figure whose area
increases with the angle $\theta$ between $\vec{H}$ and $\vec{I}$.
However the critical field $\pm H_c$ remains the same for all
angles within the experimental error ($\pm$10 Oe).} \label{rh110}
\end{figure}

In fig. \ref{rh110} the R(H)/R(0) vs. H data for the [110]
sample at 10 K (left panel) and 300 K right panel) taken at
various angles between the applied field and the current direction
is shown. The current in this case was flowing along the easy axis
\cite{Easy} and the hard axis was in-plane and 90$^0$ to the
current direction. Arrows in the figure mark the trajectory
followed by the resistance as the field was swept from positive to
negative extremities. At 10 K, the resistance increases
superlinearly as the field is reduced from 700 Oe till it reaches
a critical negative value H$_c$. On increasing the field further
in the reverse direction, the resistance drops, first rapidly and
then gradually. The resistance profile during --H$_{max}$ to
+H$_{max}$ field sweep is a mirror image of the +H$_{max}$ to
--H$_{max}$ sweep. A large hysteresis is evident in the figure
whose area increases with the angle $\theta$ between $\vec{H}$ and
$\vec{I}$. However the critical field $\pm H_c$ remains the same
for all angles within the experimental error ($\pm$10 Oe), and
also compares well with the coercive field deduced from the M-H
loop\cite{Easy}. For the measurement performed at 300 K, the R(H)
curve is mostly reversible except for a narrow range of the field
between $\pm H_c$ where twin peaks appear in the resistance for
non-zero values of $\theta$. The reversible part shows a $\rho
\propto H^\tau$ dependence with $\tau$ nearly \emph{independent}
of the angle $\theta$.

\begin{figure}
\centerline{\includegraphics[width=10cm]{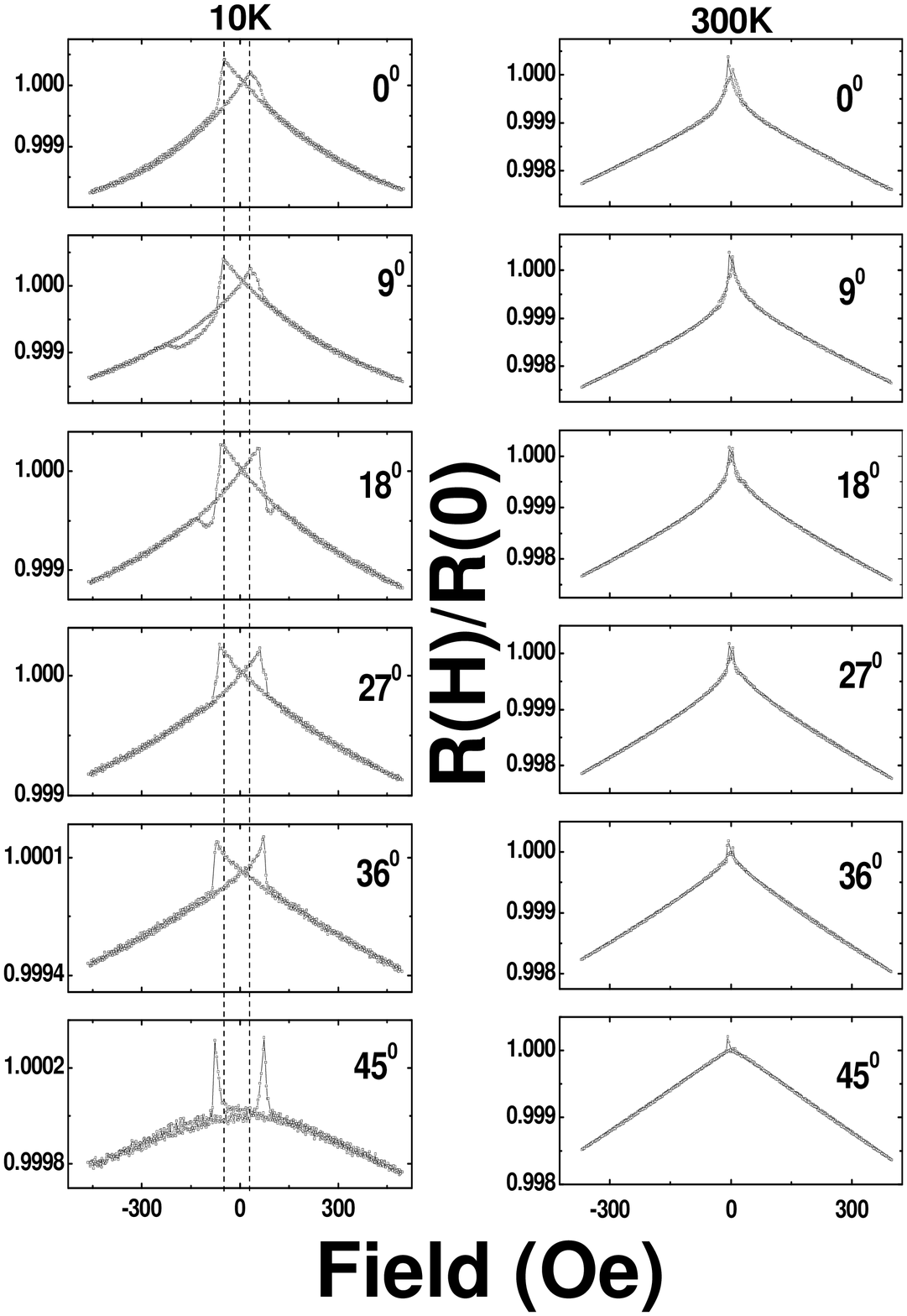}}
\caption{The isothermal magnetoresistance at different angles
between $\vec{I}$ and $\vec{H}$ for the films with [001]
orientation for measurements performed at 300 K (right panel) and
10 K (left panel). The current in this case is  flowing along the
hard axis and the easy axis is 45$^0$ to the current direction. At
10 K and $\theta = 0^0$, the R(H) curve is mostly reversible
except for the twin peaks appearing at $\pm H_c$ which agrees with
the coercive field deduced from M-H measurements. On increasing
the angle $\theta$ (moving away from the easy axis), two
interesting features emerge from the data. First, the critical
field at which resistivity drops precipitously shifts to higher
values, and second, the field dependence on increasing field
becomes superlinear to sublinear. At 300 K, $\rho(H)$ is devoid of
detectable hysteresis. Also the field dependence in this case
remains linear at all angles.} \label{rh100}
\end{figure}

The isothermal magnetoresistance at different angles between
$\vec{I}$ and $\vec{H}$ for films with [001] orientation is shown
in fig. \ref{rh100} for measurement performed at 300 K (right
panel) and 10 K (left panel). The  current in this case was
flowing along the hard axis and the easy axis \cite{Easy} was
45$^0$ to the current direction. At 10 K and $\theta = 0^0$ the
R(H) curve is mostly reversible except for the twin peaks
appearing at $\pm H_c$ which agrees with the coercive field
deduced from M-H measurements. On increasing the angle $\theta$
(moving away from the easy axis), two interesting features emerge
from the data. First, the critical field at which the resistivity
drops precipitously shifts to higher values and second, the field
dependence on increasing field becomes superlinear to sublinear
($\rho \propto H^\tau, \tau = -8.8 \times 10^{-4} \mbox{ at }
\theta = 0, \tau = -1.6 \times 10^{-4} \mbox{ at } \theta =
45^0$). At 300 K,  $\rho(H)$ is devoid of detectable hysteresis.
Also the field dependence in this case remains linear at all
angles.

The primary factors that contribute to the low field MR in these
systems are the colossal magnetoresistance effect and the
tunneling magnetoresistance if the system has a non zero
granularity. The explanation in the case of a non-granular film
lies in the multidomain configuration model proposed by O'Donnell
et al. \cite{Ecksteinprb2}.

At high fields, the magnetization is aligned in the direction of the applied field. But as the field is
lowered and then applied in the opposite direction, the
magnetization has to reverse at some point. Considering that the
reversal will be rapid, one would expect a change in resistivity
$\Delta\rho$ due to colossal magnetoresistance (CMR) given as
\cite{Ecksteinprb2}\begin{equation} \Delta\rho \approx
\rho\left(M_0 + \chi H_{sw}\right) - \rho\left(M_0 - \chi
H_{sw}\right)\label{cmrm}\end{equation} where $H_{sw}$ is the
switching field (the field at which the magnetization reversal
occurs), $M_0$ is the spontaneous magnetization and $\chi$ is the
susceptibility. This change will be seen for a transition in
magnetization from anti-parallel $\left(M \approx M_0 - \chi
H_{sw}\right)$ to parallel $\left(M \approx M_0 + \chi
H_{sw}\right)$ to the applied field. As pointed out by O'Donnell
et al. \cite{Ecksteinprb2}, this simple model cannot explain some
features in these data. Considering the case when the film is
treated as a single domain, the magnetization either flips
directly from antiparallel to parallel alignment or it comes to
the parallel alignment following a two step process, passing from
antiparallel, to transverse, to parallel. When $\vec{M}$ is
parallel or antiparallel to the applied field one
has,\begin{equation}\left|\vec{M}\right| \approx M_0 \pm \chi
H\label{singledomainm}\end{equation} and for $\vec{M}$
perpendicular to the applied field, $\left|\vec{M}\right| \approx
M_0$ which is independent of $H$. Below T$_c$ and at very low
fields, one can assume that $\chi H \ll M_0$ for a single domain
sample. So in this case the linear approximation of eq.
\ref{singledomainm} is correct. So the CMR below T$_c$ is a first
order expansion in the small parameter $\chi H$ about
$\rho\left(M_0\right)$. From these arguments and taking into
account that the CMR is linear, the change in resistivity upon
flipping of the magnetization from antiparallel to parallel
expressed by eq. \ref{cmrm} can be approximated to
\begin{equation}\Delta\rho = 2\chi H_{sw}
\left.\frac{d\rho}{dM}\right|_{M_0}\end{equation} The
single-domain model also fails to explain the deviation from
linearity. To overcome the shortcomings of the single domain
model, O'Donnell et al. \cite{Ecksteinprb2} proposed a
multidomain-model. In this model, the magnetization reversal
proceeds via motion of the domain walls. The resistivity for a
sample with the applied field along the easy axis can be written
as \cite{Ecksteinprb2}
\begin{equation}\rho \approx x\rho_{par} + y\rho_{antipar} + z\rho_{transverse}\label{multidomain}\end{equation} where $\rho_{par},
\rho_{antipar} \hbox{ and } \rho_{transverse}$ are the
resistivities of the domains parallel,  antiparallel and
transverse to the applied field.

Now, the MR data is analyzed in the light of the multi-domain
model taking into account that [110] LSMO films show uniaxial and
[001] films show biaxial anisotropy. Looking at the data for [110]
samples at 300 K (right hand panel of fig. \ref{rh110}), no
discontinuous change in resistivity when $H \| I$ is seen. So it
looks like that as soon as the field is reversed, the
magnetization also reverses. For other angles, a clear
discontinuous change is seen. This behavior can be understood by
arguing that as the field is slowly increased from zero, the
moments align along the hard axis that is perpendicular to the
current and as a result, the resistivity increases. Once the field
crosses a threshold limit, the magnetization flips abruptly
towards the applied field direction. Since T=300 K is very close
to $T_c$ for LSMO, it is very weakly ferromagnetic at that
temperature. Hence it is possible that for fields aligned along
the easy axis no discontinuous jump may be seen. The data at 10 K
(left hand panel of fig. \ref{rh110}) show large hysteretic loop
in MR. Secondly, the transition is not very sharp. This can be
attributed to the deviation from the square hysteresis loop that
this sample shows at 10 K\cite{Easy}. In this case, a transition
is also seen for $H\| I$ since at 10 K, LSMO is strongly
ferromagnetic.

The R(H)/R(0) curves for the [001] sample reveal some interesting
facts as well. If one looks at the data at T = 300 K (right hand
panel of fig. \ref{rh100}), a linear R-H behavior when $H$ is
parallel to the easy axis (i.e. $45^0$ to the current direction)
\cite{Easy} is seen.  Some non-linearity sets in once the applied
field direction moves away from the easy axis but the
non-linearity is not hysteretic. The observed non-linear
dependence presumably arises from the slow rotation of the
magnetization vector with the increasing external field.

The data at 10 K (left hand panel of fig. \ref{rh100}) show some
features of first order transition but the explanation is a little
different in this case since [001] samples show biaxial
anisotropy. So, when the applied field is along the
current direction at low fields, domains aligned in the direction
of the easy axis start appearing. Since these domains are aligned
away from the current direction the resistance increases. At a
particular field, the moments switch to the direction of the
applied field and one sees the step in the resistivity. But in
this case, the step is also present for $H$ parallel to the easy axis
that is 45$^0$ to the current direction. This can be attributed to
the fact that for the field to flip completely, it has to cross
two hard directions. So a minimum magnetic energy is required to
completely flip the moments.

\section{Conclusion}
In conclusion it is shown that the anisotropy energy of [110]
La$_{2 / 3}$Sr$_{1 / 3}$MnO$_{3}$ films is larger than that of
[001] La$_{2 / 3}$Sr$_{1 / 3}$MnO$_{3}$ films. Further [110] films
show larger hysteresis than that shown by [001] films. The data
has been analyzed in the light of multidomain model and the
behavior seen in [001] and [110] samples can be explained
correctly by this model. It is also shown, for both the films the
resistivity in the reversible regime is proportional to $H^\tau$,
with $\tau$ nearly independent of the angle of the applied field
for [110] film which is not the case for [001] film.

The author would like to acknowledge financial support from Indian
Institute of Technology Kanpur, Kanpur, India.

\end{document}